\documentclass[fleqn,usenatbib]{mnras}

\usepackage[T1]{fontenc}

\DeclareRobustCommand{\VAN}[3]{#2}
\let\VANthebibliography\thebibliography
\def\thebibliography{\DeclareRobustCommand{\VAN}[3]{##3}\VANthebibliography}

\usepackage{graphicx}	% Including figure files
\usepackage{amsmath}	% Advanced maths commands
\usepackage{amssymb}	% Extra maths symbols
\usepackage{layouts}
\usepackage[x11names]{xcolor}
\newcommand{\astronomaly}{\textsc{astronomaly}}
\newcommand{\pybdsf}{\textsc{pybdsf}}
\newcommand{\opencv}{\textsc{opencv}}
\newcommand{\astropy}{\textsc{astropy}}

\newcommand{\review}[1]{{\textcolor{black}{#1}}}
\title[A unique ring-like radio source]{A unique, ring-like radio source with quadrilateral structure detected with machine learning}

\author[M. Lochner et al.]{
M. Lochner,$^{1,2}$\thanks{E-mail: dr.michelle.lochner@gmail.com}
L. Rudnick,$^{3}$
I. Heywood,$^{4,5,2}$
K. Knowles$^{5,2}$
and S. S. Shabala$^{6}$
\\
$^{1}$Department of Physics and Astronomy, University of the Western Cape, Bellville, Cape Town, 7535, South Africa\\
$^{2}$South African Radio Astronomy Observatory, 2 Fir Street, Black River Park, Observatory, 7925, South Africa\\
$^{3}$Minnesota Institute for Astrophysics, University of Minnesota, 116 Church St SE, Minneapolis, MN 55455, USA\\
$^{4}$Astrophysics, University of Oxford, Denys Wilkinson Building, Keble Road, Oxford, OX1 3RH, UK\\
$^{5}$Centre for Radio Astronomy Techniques and Technologies, Department of Physics and Electronics, Rhodes University, \\PO Box 94, Makhanda 6140, South Africa\\ 
$^{6}$School of Natural Sciences, Private Bag 37, University of Tasmania, Hobart, TAS 7001, Australia\\
}

\date{Accepted XXX. Received YYY; in original form ZZZ}

\pubyear{2022}

\usepackage{newtxtext,newtxmath}
\begin{document}
\label{firstpage}
\pagerange{\pageref{firstpage}--\pageref{lastpage}}
\maketitle

\begin{abstract}
We report the discovery of a unique object in the MeerKAT Galaxy Cluster Legacy Survey (MGCLS) using the machine learning anomaly detection framework \astronomaly{}. This strange, ring-like source is 30\arcmin{} from the MGCLS field centred on Abell 209, and is not readily explained by simple physical models.  With an assumed  host galaxy at redshift 0.55,  the luminosity ($10^{25}$ W Hz$^{-1}$) is comparable to powerful radio galaxies. The source consists of a ring of emission 175 kpc across,  quadrilateral  enhanced brightness regions bearing resemblance to radio jets, two ``ears'' separated by 368 kpc, and a diffuse envelope.  All of the structures appear spectrally steep, ranging from -1.0 to -1.5.  The ring has high polarization (25\%) except on the bright patches ($<10$\%). We compare this source to the Odd Radio Circles recently discovered in ASKAP data and discuss several possible physical models, including a termination shock from starburst activity, an end-on radio galaxy, and a supermassive black hole merger event. No simple model can easily explain the observed structure of the source. This work, as well as other recent discoveries, demonstrates the power of unsupervised machine learning in mining large datasets for scientifically interesting sources.
\end{abstract}

\begin{keywords}
galaxies:active -- radio continuum:galaxies
\end{keywords}

%%%%%%%%%%%%%%%%%%%%%%%%%%%%%%%%%%%%%%%%%%%%%%%%%%

%%%%%%%%%%%%%%%%% BODY OF PAPER %%%%%%%%%%%%%%%%%%

\section{Introduction}
% \thefontsize{}\\
% \printinunitsof{pt}\prntlen{\columnwidth}\\
Over a billion radio sources are expected to be catalogued by surveys with the Square Kilometre Array \citep[SKA,][]{SKA}. Already, the SKA precursor telescopes, such as MeerKAT \citep{MeerKAT}, ASKAP \citep{ASKAP} and LOFAR \citep{LOFAR}, are producing datasets of remarkable size and richness. These datasets have proven to be a treasure trove for scientific discoveries, including Odd Radio Circles \citep[ORCs][]{Norris2021}, the ``heartworm'' nebula \citep{Cotton2022}, a giant 5-Mpc radio galaxy \citep{Oei2022} and many others. Even older large surveys, such as FIRST \citep{Becker1995}, have continued to reveal interesting new sources through intensive visible inspection in projects such as Radio Galaxy Zoo \citep{Banfield2015,Banfield2016}. However, these discoveries were made by scientists and citizen scientists serendipitously noticing the unusual source and the increasing volume of modern datasets means that many interesting sources may be missed. An alternative to manual search is machine learning: a set of algorithms designed to automatically learn patterns and models from data, which has the potential to assist in rapidly sorting through data to locate interesting sources in large astronomical datasets. Recent unsupervised machine learning approaches have proven very effective at discovering radio galaxies with unusual morphology \citep{Segal2019, Segal2022, Galvin2020, Mostert2021, Gupta2022}.

\astronomaly{} \citep{Lochner2021} is a general machine learning framework for anomaly detection in astronomical data. It uses a novel active learning approach to optimally combine blind anomaly detection algorithms with a small amount of human labelling to quickly discard uninteresting anomalies, such as artefacts. This allows a scientist to focus most of their attention on a much smaller subset of interesting sources in the data. \astronomaly{} has been used to detect optical galaxies with unusual morphology \citep{Walmsley2022} and to discover anomalous optical transients \citep{Webb2020}. 

In this work, we used \astronomaly{} to search for sources with unusual morphology in the MeerKAT Galaxy Cluster Legacy Survey \citep[][MGCLS]{Knowles2022}. MGCLS is a programme of MeerKAT L-band (900--1670 MHz) observations of 115 galaxy clusters. The full MGCLS survey contains $\sim 720,000$ sources, with an expected $\sim 6,000$ multi-component extended sources (extrapolated from extended source counts in two catalogued fields). While the size of this dataset may not seem particularly large, in practice it is an extremely \emph{rich} dataset and many interesting sources have not yet been studied. During this search, \astronomaly{} detected a source with unique morphology which was missed by scientists and is the subject of this paper. 

In \autoref{sec:methodology} we briefly describe the process by which the source was discovered, \autoref{sec:data} details reprocessing of the original MGCLS data. We present an analysis of the properties of the source in \autoref{sec:analysis}, discuss possible models in \autoref{sec:models} and conclude in \autoref{sec:conclusions}. We assume cosmological parameters from \cite{Planck} throughout the paper.

\section{Methodology}
\label{sec:methodology}
To detect interesting sources in the 115 images, covering 166 square degrees, from the enhanced MGCLS data products, we first ran the \pybdsf{} source-finding algorithm \citep{Mohan2015} with standard parameters. \review{\pybdsf{} works by applying a threshold 
dependent on the local rms noise,
and then detecting regions of contiguous emission, called islands. The islands are fit with a number of Gaussian components, which are usually used in radio source catalogues. However, because our focus is on sources with unusual morphology, we obtained better results by constructing the catalogue from \pybdsf{} islands rather than individual Gaussian components.} We removed the majority of simple, uninteresting sources by requiring a minimum of four components per island. We then used the average position between the components as the source centre, and also determined the minimum size for the source using the island boundary. The final image for each source is thus scaled to ensure all emission is captured in a single cutout. We resized each image, interpolating if necessary, to 128x128 pixels. 

\review{We applied a standard sigma-clipping algorithm from \astropy{} \citep{astropy:2013, astropy:2018, astropy:2022}, using a threshold of $3\sigma$. This algorithm uses an iterative approach to estimate the noise of the image and then masks pixels below the $3\sigma$ threshold. This selects the brightest regions in the cutout but includes unrelated background sources that may confuse the machine learning algorithm. To correct for this, we applied a contour-fitting algorithm from \opencv{} \citep{Bradski2000} to the masked image and selected only the emission contained in the central contour (assumed the be the source in question) thus removing the background sources.} This approach resulted in a catalogue of 6332 extended sources. 

After extracting cutouts of the sources, we followed a similar procedure as in \cite{Lochner2021} which works well for radio data despite originally being applied to optical galaxies. The approach combines simple morphological features with the anomaly detection algorithm isolation forest \citep{liu2008} to provide an initial ranking according to anomaly score. We note that due to technical failures in the feature extraction (usually for small sources with few data points), the total number of sources reduced to 6047.

\astronomaly{} ranks each object in the dataset according to anomaly score and presents these in a visual interface. We manually evaluated the 200 most anomalous sources, giving them a score out of 5 for ``interestingness''. This is inherently subjective, but we focused on radio galaxies with complex and unusual morphology. Finally, we applied the novel active learning algorithm of \cite{Lochner2021} and sorted the data by the active anomaly score. Through this process, we discovered a source with very unusual morphology, which is the topic of this paper. We note that this source was ranked in the top 1.2\% of the unlabelled data after applying active learning, illustrating how rapidly \astronomaly{} can discover interesting sources with only a very small amount of human labelling.

\review{Located at right ascension and declination 22$^{\circ}$.9895, -13$^{\circ}$.5763},
the source discovered by \astronomaly{} has a ring-like structure with four, uneven hot spots and ears extending to either side. It resembles an Odd Radio Circle \citep[ORC,][]{Norris2021} and in deference to the ORC theme, we will refer to it  as SAURON - a Steep And Uneven Ring Of Nonthermal radiation.

\section{Reprocessing of the Abell 209 field data}
\label{sec:data}
The enhanced MGCLS data products for the Abell~209 field include Stokes I, Q and U cubes, split into 14 frequency bins across 856-1712~MHz. See \cite{Knowles2022} for details on the observations and data processing. The images are centred at right ascension and declination 22$^{\circ}$.9895, -13$^{\circ}$.5763 and focus on the inner $1.2^{\circ}\times1.2^{\circ}$ of the pointing. While we made use of the original Stokes Q and U cubes in our analysis, we used a reprocessed version of the Stokes I data to obtain a clearer view of SAURON.

We reprocessed the Abell 209 data (block ID 1538591458, 3 October 2018) following the workflow \citep[{\sc oxkat} v0.3;][]{heywood2020} that was implemented for the continuum imaging of the MIGHTEE survey \citep{jarvis2016} and described in more detail in \citet{heywood2022}. Briefly, the {\sc casa} package  \citep{mcmullin2007} was used to apply flags to and generate calibration solutions from the primary (PKS 0408$-$65) and secondary (J0059+0006) calibrators. The solutions were applied to the target data, which were then flagged using the {\sc tricolour} package \citep{hugo2022}. Imaging and delay self-calibration was performed using {\sc wsclean} \citep{offringa2014} and {\sc cubical} \citep{kenyon2018}, and facet-based direction-dependent corrections were applied using {\sc ddfacet} \citep{tasse2018} and {\sc killms} \citep{smirnov2015}.

\begin{figure}
 \centering
  \includegraphics[width=1\linewidth]{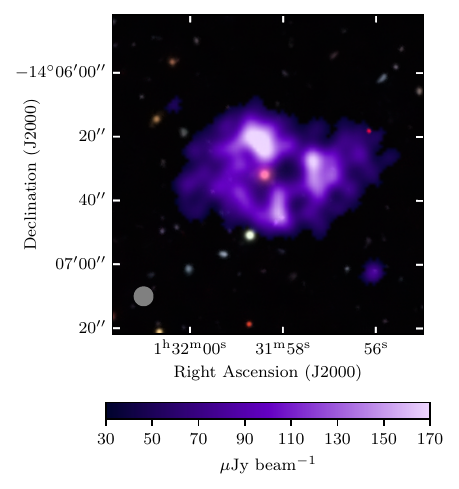}
  \caption{Radio intensity (purple) overlaid on an optical image from the Dark Energy Survey \citep{DES, Abbott2021, Morganson2018} indicating a possible centrally located counterpart at $z$=0.557. \review{We set all pixels in the radio overlay below $3\sigma$ to zero for cosmetic purposes, to ensure noise does not obscure the optical background. The colour bar is restricted to the range $30-170\,\mu\rm{Jy}\, \rm{beam}^{-1}$ to enhance the contrast. The synthesised beam (FWHM 6.2\arcsec{}) is shown in the lower left corner.} }
\label{fig:optical}
\end{figure}

Sub-band images for spectral indices were made in 11 frequency chunks of equal fractional bandwidth. A \citet{briggs1995} \review{robustness} of $-$0.4 was used with a consistent Gaussian taper, and a tapered cosine window of width 400 wavelengths applied to the inner part of the $u$,$v$ plane. The model images were convolved with a circular 10$\farcs$5 Gaussian. The residual images were convolved with a homogenisation kernel determined using the {\sc pypher} package \citep{boucaud2016}, and summed with the convolved model. Primary beam correction was performed using the {\sc katbeam}\footnote{\url{https://github.com/ska-sa/katbeam}} package.

\autoref{fig:optical} shows the final total intensity radio image overlaid on optical data. The \emph{griz} DES DR1 images were combined into an RGB image using a preprocessing function from \astronomaly{}\footnote{\url{https://github.com/MichelleLochner/astronomaly/blob/main/astronomaly/preprocessing/image_preprocessing.py\#L365}} which rescales each band by multiplying by a given scale factor ($g$:6.0, $r$:3.4, $i$:1.0, $z$:2.2) and applying an arcsinh stretch. The $i$ and $z$ bands were combined to form the R plane, the $r$ band was used for G and the $g$ band for B. A Non-Local Means Denoising algorithm \citep{Buades2011} was applied using \opencv{} (with $h=17$) to remove a small amount of background noise and obtain a clearer picture.

\section{Analysis of SAURON}
\label{sec:analysis}
 
In order to understand possible mechanisms that may produce this source, we studied the morphology, spectral indices and polarization properties of SAURON, as well as any multiwavelength data available.
 
The most \review{likely} optical counterpart, seen at the centre of SAURON in \autoref{fig:optical}, is WISEA~J013158.40-140631 \citep{WISE}. This galaxy has a photometric redshift of $0.557\pm0.008$, according to the DES Y3 GOLD catalogue \citep{DrlicaWagner2018,SevillaNoarbe2021}. The infrared and photometric colours from WISE and DES meet the criteria laid out in \citet{Zhou2022}, making the host a likely Luminous Red Galaxy. There are no other galaxies in the DES catalogue within 10\arcsec{} of the centre of the source. 
\begin{figure}
 \centering
  \includegraphics[width=1\linewidth]{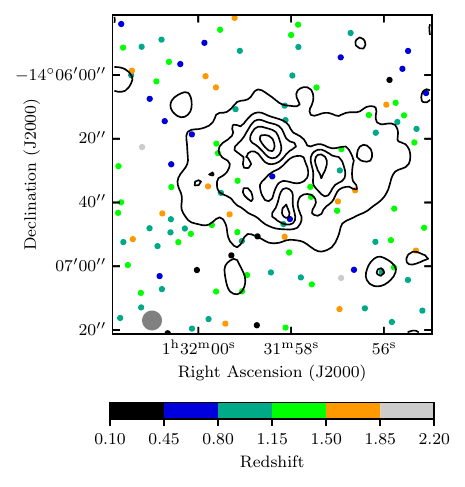}
  \caption{\review{Total intensity  contours at $(3, 10, 15, 20, 25) \times$ the rms value of 7.3$\mu \rm{Jy}\, \rm{beam}^{-1}$ with all galaxies in the region from the DECaLS DR2 catalogue plotted in colours corresponding to their photometric redshift estimate. The synthesised beam (FWHM 6.2\arcsec{}) is shown in the lower left corner.}}
\label{fig:des_sources}
\end{figure}

\review{\autoref{fig:des_sources} shows all detected optical sources from the DECaLS DR8 catalogue \citep{decals} in the vicinity of SAURON, coloured by their photometric redshift\footnote{\review{We use DECaLS for this particular figure because the catalogue has been produced with fewer cuts than the DES catalogue. So while this will include likely false positive sources, it is the most conservative test to look for any possible counterparts for the hot spots.}}. While several galaxies are detected
within the contours of the radio emission,
they do not align convincingly with the radio peaks 
and are all found at different redshifts, suggesting they are unrelated background or foreground sources. The lack of optical counterparts for the hot spots
and the excellent alignment of all hot spots with the ring makes it highly unlikely, but not impossible, that the hot spots and the ring are not connected to the same host galaxy and are simply a coincident alignment of unrelated sources.}

The area around SAURON has a density of 1522 galaxies per square degree (r\textless19.81)\footnote{Using the DES DR2 MAIN table and selecting all objects with quality flags less than 4 and EXTENDED\_CLASS\_COADD\textgreater2 in a square degree box around SAURON.}, meaning there is only a 3.7\% probability of coincidentally detecting a source of this brightness in a 10\arcsec{} radius around the centre of the source.  For the rest of this paper, all size and distance measurements are made under the assumption that this is the correct host galaxy and using the cosmological parameters from \citep{Planck}. Unlike other ORCs with central hosts, we find no evidence of radio emission, with a 3$\sigma$ upper limit of 30~$\mu$Jy, or $<0.8\%$ fractional luminosity.

The source does not appear to be in any known cluster or group of galaxies. Quantifying the true density of the environment is difficult without a large sample of spectroscopic redshifts. However, the photometric redshift distributions from the DES Y3 GOLD sample indicate a $7.2\pm2.5$\% excess of galaxies in the redshift range $0.5-0.6$ in a $10\arcmin{}$ radius region around SAURON compared to the background (averaged over 500 independent regions of similar size). It thus does appear that SAURON may be in a slightly overdense region. 

\begin{figure*}
 \centering
  \includegraphics[width=1\linewidth]{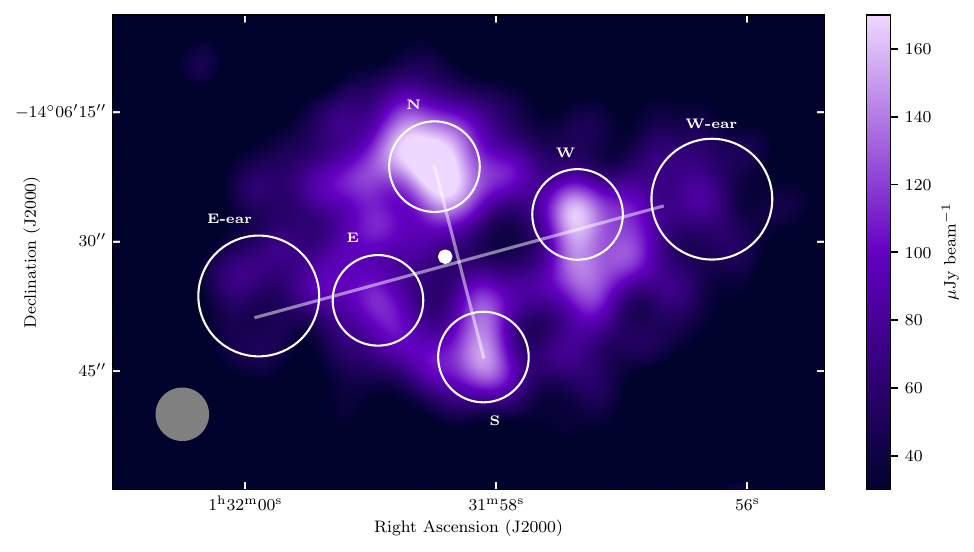}
  \caption{Total radio intensity (purple) with annotations overplotted indicating the four hot spots (N, S, E and W), the two ears (E-ear and W-ear), the possible host galaxy (dot in the centre) and two perpendicular lines suggesting quadrilateral symmetry between the regions of interest. The size of the regions also indicate the areas used to compute the spectral index in \autoref{fig:spectrum}, which are 10.5\arcsec{} diameter for the hot spots and 14\arcsec{} for the ears. \review{The colour bar is restricted to the range $30-170\,\mu\rm{Jy}\, \rm{beam}^{-1}$ to enhance the contrast. The synthesised beam (FWHM 6.2\arcsec{}) is shown in the lower left corner.}}
  \label{fig:annotations}
\end{figure*}

The regions of interest we refer to throughout the paper are shown in \autoref{fig:annotations}. SAURON has four ``hot spots'', with the \review{east} hot spot being considerably fainter than the others. It also has two ears embedded in the envelope. Its structure suggests quadrilateral symmetry, as indicated by two perpendicular lines - one connecting the \review{north} and \review{south} hot spots, the other bisecting the first and at a $90^\circ$ angle. While this symmetry is not perfect, it is compelling, and any physical model must be able to explain both the ring and the structure of the hot spots and ears. At the \review{redshift of the assumed host (0.557)}, SAURON's  ring is 175 kpc in diameter; the ears and broader envelope are at least 370~kpc in extent. Its peak \review{brightness} is 0.2 mJy beam$^{-1}$ with an integrated flux \review{density} of 3.9 mJy. This corresponds to a monochromatic 1.4~GHz rest frame luminosity of $\sim10^{25}$ W Hz$^{-1}$, in the range of powerful radio galaxies.

\begin{figure}
 \centering
  \includegraphics[width=0.99\columnwidth]{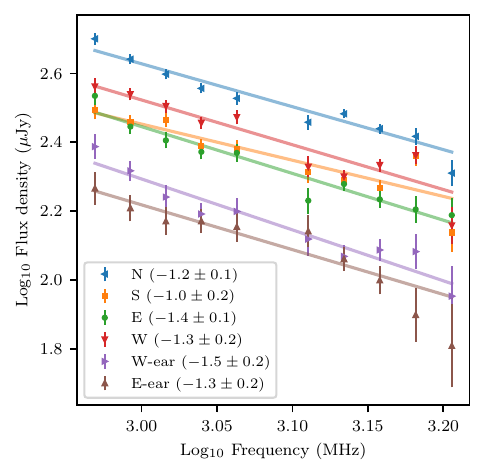}
  \caption{Spectrum of various regions (defined in \autoref{fig:annotations}) of  SAURON. The spectral index of each region is given in the legend with its error.}
  \label{fig:spectrum}
\end{figure}

\begin{figure}
 \centering
  \includegraphics[width=0.99\columnwidth]{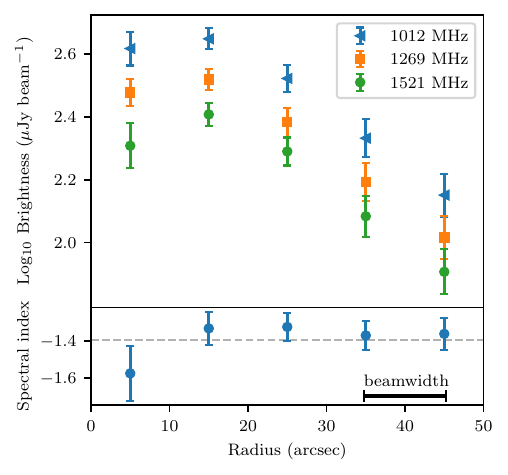}
  \caption{\emph{Top} - radial profiles of the brightness averaged over three sets of frequency channels, labeled with the central frequencies in each set. \emph{Bottom} - the average spectral index over all frequency channels.}
  \label{fig:radial_spectrum}
\end{figure}

\autoref{fig:spectrum} shows plots of the spectra for the features defined in \autoref{fig:annotations}.  The spectral indices are estimated using a simple weighted least squares method. \review{We average over an aperture of diameter 10.5\arcsec{} for the hot spots and 14\arcsec{} for the ears.} The higher frequency bands contain some visually spurious points which may result from residual radio frequency interference, and  may bias the spectral index estimates at a level of $\pm$0.2. More robust spectral indices from on-axis observations of SAURON would be useful to quantify the variations between individual features. \review{The fact that the hot spots all have a similar spectral index to each other and the ring, and that spectral index ($\sim-1.2$) is quite different from the average spectral index of galaxies in the MGCLS catalogue ($\sim-0.7$) is further evidence that the hot spots and ring form part of the same source and are not coincident alignments.}
\autoref{fig:radial_spectrum} shows the spectra calculated by averaging the flux \review{density} in 10\arcsec{} rings from the centre of the source. 
There is a clear drop in flux \review{density} in the centre of the source. The spectrum appears relatively constant with radius, with marginal evidence for a steeper spectrum in the centre.

\begin{figure}
 \centering
  \includegraphics[width=0.99\columnwidth]{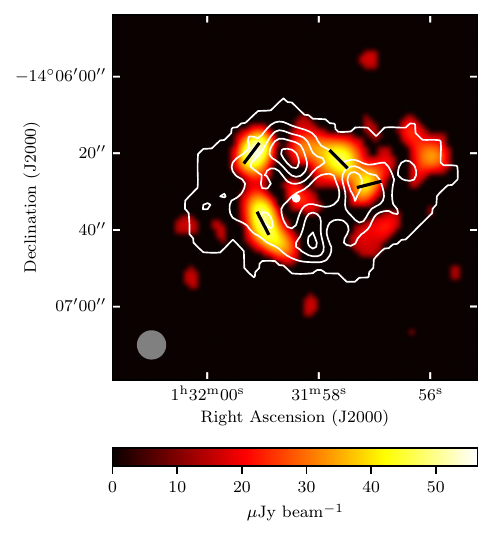}
  \caption{\review{Average} bias corrected polarized flux \review{density, over the range $908-1145$ MHz} in orange and white total intensity  contours at $(3, 10, 15, 20, 25) \times$ the rms value of 7.3$\mu \rm{Jy}\, \rm{beam}^{-1}$. \review{Black vectors indicate the inferred magnetic field directions at the positions of the four peaks in polarized intensity, after correction for the local RM.}  The white dot at the centre represents the potential host galaxy and \review{background emission is removed with sigma clipping at a $\sim3\sigma$ level for the total intensity contours}. \review{The synthesised beam (FWHM 7.6\arcsec{}) is shown in the lower left corner.}}
  \label{fig:polarisation}
\end{figure}
 
\autoref{fig:polarisation} shows the source polarized flux \review{density}, corrected for the noise bias, \review{using only the range $908-1145$ MHz where the signal:noise was highest}. The vectors indicate the inferred magnetic field direction, after correcting for rotation measure (RM), and have uncertainties of $\sim$10$^{\circ}$.  The RMs were derived from least-squares fits to Q and U at the peaks of polarized emission \review{for the entire range of frequencies}, and gave values in the range $3-8~\rm{rad}~\rm{m}^{-2}$. \review{Characteristic errors were 
$\sim\pm5-10~\rm{rad}~\rm{m}^{-2}$. The distribution of residuals is non-Gaussian, so the error estimate depends on the choice of how spurious individual values were excluded from the calculation.} The Galactic foreground is $\sim1\pm5~\rm{rad}~\rm{m}^{-2}$ in this direction,\footnote{Using the CIRADA RM cutout server (\url{http://cutouts.cirada.ca/rmcutout/}) based on \cite{Hutschenreuter2022}.} so there is no strong Faraday contribution local to SAURON.
Away from the hot spots, the fractional polarizations are $\sim$25\%, similar to that seen for ORC1 \cite{Norris2022}. However, there is no evidence for polarized emission associated with the hot spots.  The observed fractional polarizations of 3-5\% appear to contain contributions from the surrounding polarized ring emission.  Conservatively, we can set an upper limit of $\sim$10\% on the hot spot polarized fractions, but the actual value is likely considerably smaller. The fact that the regions of high polarization do not align with the hot spots implies they are not simply enhanced regions of the ring and rather arise from a different physical process.

There are no sensitive public X-ray data available for SAURON.
However, a brief investigation of \textit{SRG's eRosita} \citep{ERosita} data in this region revealed no X-ray emission from the position of the host, at an approximate sensitivity of $10^{-13}~\rm{erg}~\rm{s}^{-1}~\rm{cm}^{-2}$ (Churazov, E., Sunyaev,S., Khabibullin,I., private communication), corresponding to an approximate luminosity of $<10^{45}~\rm{erg}~\rm{s}^{-1}$. Thus, there is not a high luminosity AGN present, although more sensitive observations would be needed to rule out less luminous AGN activity \citep{Fotop16}. 

\begin{figure}
  \centering
   \includegraphics[width=0.99\linewidth, trim={0 0 0 0pt}, clip]{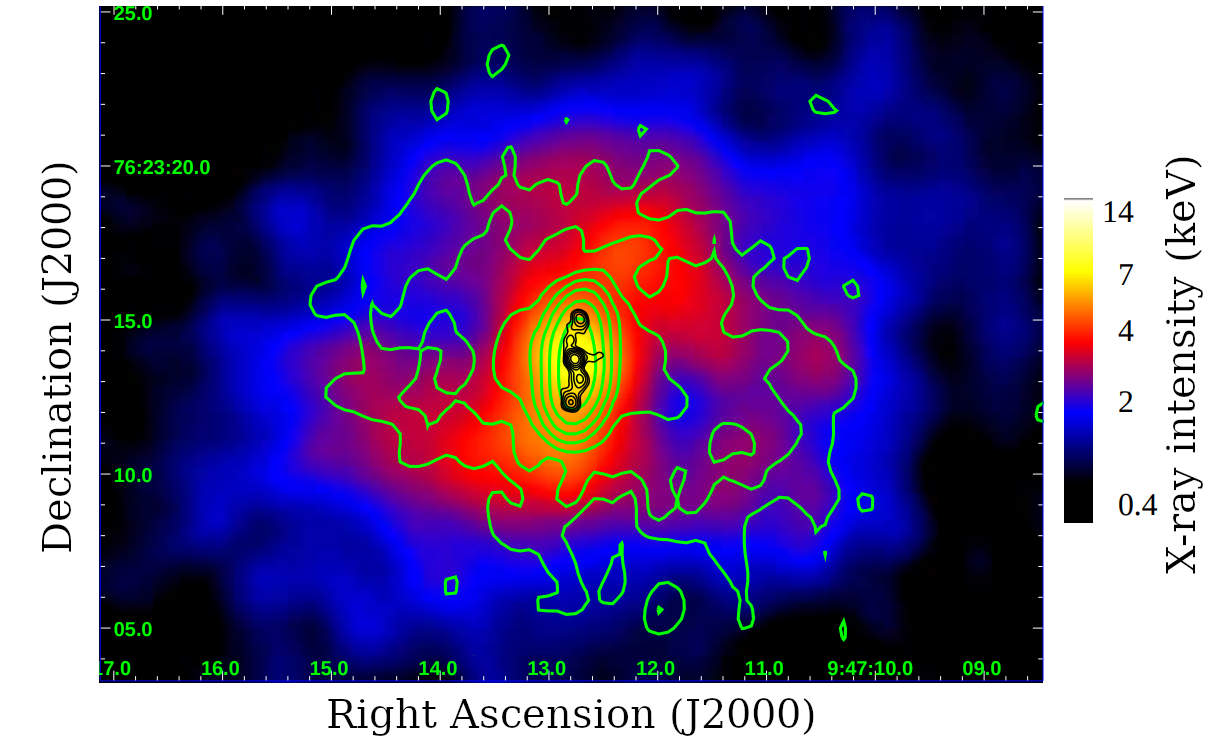}
   \caption{1.4GHz (green) and 4.8GHz (black) VLA radio contours overlaid onto the Chandra X-ray image for the possible binary black hole in RBS~797. Credit: \citeauthor{Gitti2006}, A\&A, 448, 853–860, 2006, reproduced with permission © ESO.}
   \label{fig:xray}
\end{figure}

 \section{ORC physical models}
 \label{sec:models}
 In many ways, SAURON is similar to the ORCs with central hosts described by \cite{Norris2022}.  Despite their very different morphological appearances, the similarities between ORC1 and SAURON are striking, as seen in Table 1.  Thus, much of the detailed analysis of physical models by \cite{Norris2022} applies to SAURON as well. Some derived parameters would change, e.g., SAURON has the same luminosity but a smaller emitting volume, so the magnetic fields are likely somewhat higher. But the overall conclusions remain --   the sizes, luminosities, energetics, radio spectra, radiative timescales, etc.  can all be plausibly explained by several different ring-producing mechanisms. These include a termination shock from intense starbursting activity, an explosion following the merger of SMBHs, and the end-on view of a radio galaxy.  Each of these possible mechanisms also had significant flaws, for example, the unexplained interior arcs of ORC1, so the models are currently incomplete.
 \begin{table}
     \centering
     \begin{tabular}{|c|c|c|}
       Property   & ORC1 & SAURON \\
       \hline
       \hline
       Integrated flux \review{density} (mJy)   & 3.9  & 3.9\\Redshift  & 0.55 & 0.55 \\
       Spectral index & -1.4$^a$ & -1.4 \\
       Ring \% polariz. & 15-30 & 10-28 \\
       Ring/\review{Interior} brightness$^b$ & 1.7 & 1.6 \\
       Diameter (kpc) &520 &{\em 370 (175)$^c$} \\
        \% host flux \review{density}  & 2.7 & {\em $<$0.8}\\
       WISE W1 (mag) & 15.065 & 15.058\\
       W1-W2 (mag) & 0.081 & 0.001 \\
       W2-W3  (mag) & $<$2.1 & $<$2.6 \\
       DES g-r (mag)& 1.9 & 1.79 \\
       DES r-i (mag) & 0.87 & 0.82 \\
       \hline
     \end{tabular}
     \caption{Comparison of SAURON and ORC1 (values taken from \citet{Norris2021, Norris2022}). The only two values that are significantly different between the two sources are noted in italics. Notes: a. Multi-telescope spectral index; b. Average radio brightness of ring / central emission, reanalysed here for ORC1. c. Full diameter,  (ring diameter).}
     \label{tab:comparee}
 \end{table}
 
SAURON, too, creates difficulties for these simplest scenarios.  Its host has no detectable radio emission, so support for either AGN or starburst activity is lacking. 

The N-S structures in SAURON are very similar in appearance to radio jets/lobes (\autoref{fig:optical}), and have significantly lower fractional polarizations than the rest of the ring, suggesting they may have formed differently.   Although ORC1 contains some relatively low fractional polarizations on its bright polarized ring, we re-examined those images and found that these low polarization regions did not correspond to ORC1's brightest patches as they do in SAURON. 

Orthogonal to the N-S jets, we find more amorphous brightness enhancements, including the ears.  This overall quadrilateral structure is reminiscent of a source in the RBS~797 cluster (\autoref{fig:xray}), a likely pair of twin radio jets forming and filling four X-ray cavities  \citep{Ubertosi2021}. The authors of that study propose that the RBS~797 source is a merging SMBH system, with VLBI information suggesting that the two sets of jets represent either a re-orientation of the central ejection axis or separate ejections from the two SMBHs.  \review{ RBS~797} has one set of well-collimated jets (coincidentally, N-S) while the orthogonal ones are more diffuse. In SAURON, the situation could be similar, e.g., with the ears representing an earlier phase of jet activity, \review{and therefore weaker than the N-S jets.}  \review{RBS~797's} radio spectral indices are similar to those of SAURON (in the range of $-1.1$ to $-1.3$), and they have the same monochromatic radio luminosities, $\sim10^{25}$~W/Hz~\citep{Gitti2006}. \review{One important difference is that RBS~797} is on a significantly smaller physical scale (roughly 30~kpc) in diameter.

On larger scales, \citet{Chon2012} detected X-ray cavities in the well-studied radio source, Cygnus A. They suggested that this was evidence of previous AGN activity. \citet{Krause2019} went further to suggest Cygnus A shows strong evidence of precession and argue, by studying several well-resolved radio galaxies, that sub-parsec supermassive binaries may be common in powerful radio sources. It is interesting to note that the model for merging SMBHs developed by \citet{Merritt2002} predicts an extremely rapid 90$^{\circ}$ spin flip of the central black hole. While this may not be a general result, it could naturally explain the unique quadrilateral structure of SAURON.

\review{Recently, \citet{Dolag2022} used advanced simulations to demonstrate that ORC-like structures could  emerge naturally as a result of shocks produced by extreme galactic merger events which result in a galaxy with a virial mass of $10^{13}~\rm{M}_{\rm{sun}}$. An alternative explanation, suggested by \citet{Omar2022}, is that ORCs could be produced by shocks from $10^5 - 10^9$  tidal disruption events around the central massive black hole  in a post-starburst galaxy.}

Whether or not the merging SMBH and two pairs of jets scenario holds, we are left with the challenge of explaining both the existence of the ring, and the quadrilateral structure, including the ears, for SAURON.  None of the simple scenarios are sufficient, similar to the situation for ORC1, although one or more is likely to be playing a role. 

\section{Conclusions}
\label{sec:conclusions}
The unusual source we have called SAURON is the first (to our knowledge) scientific discovery made with machine learning in MeerKAT image data. The anomaly detection framework, \astronomaly{}, easily detected SAURON, although it had not been previously spotted by the many people working on the MGCLS. SAURON shares many properties with ORCs, particularly ORC1.  SAURON  is thus likely explained by one or more of the physical models explored for the ORCs, including a termination shock from intense star-bursting activity, the end-on view of a radio galaxy and an explosion following the merger of SMBHs. Notably though, SAURON demonstrates unusual quadrilateral morphology with low polarization in the brightest regions,  which differs from ORC1. These results, and the fact that there is no detectable radio emission from the likely host galaxy, disfavours the star-burst and end-on radio galaxy scenarios, but does not rule them out. The quadrilateral structure of SAURON  seems supportive of a merging SMBH origin and resembles another such system in RBS~797; however the SMBH model does not explain the observed ring or ears without additional complexity.

Radio observations with SAURON on axis, and also at lower frequencies, may reveal a clearer picture of the morphology and polarization structure, while deep X-ray observations could provide support for the SMBH scenario. Additionally, detailed hydrodynamic simulations using the physical models discussed are needed to fully explain SAURON and other similar ORCs. We expect to discover many more anomalous sources in radio data using techniques such as \astronomaly{}, which will likely continue to push our understanding of radio galaxies and the physics of extreme energetic events.

\section*{Acknowledgements}
The authors thank the reviewer and editor for their helpful comments.

The authors would like to thank Eugene Churazov, Rashid Sunyaev and Ildar Khabibullin for their quick analysis of eRosita data. We would also like to thank John Kenyon for useful discussions.

ML and KK acknowledge support from the South African Radio Astronomy Observatory and the National Research Foundation (NRF) towards this research. Opinions expressed and conclusions arrived at, are those of the authors and are not necessarily to be attributed to the NRF.

%Meerkat
MGCLS data products were provided by the South African Radio Astronomy Observatory and the MGCLS team and were derived from observations with the MeerKAT radio telescope. Access to the variety of MGCLS data products is described in \cite{Knowles2022}. The MeerKAT telescope is operated by the South African Radio Astronomy Observatory, which is a facility of the National Research Foundation, an agency of the Department of Science and Innovation.

%%%WISE
This publication makes use of data products from the Wide-field Infrared Survey Explorer, which is a joint project of the University of California, Los Angeles, and the Jet Propulsion Laboratory/California Institute of Technology, funded by the National Aeronautics and Space Administration.

%%%DES
This project used public archival data from the Dark Energy Survey (DES). Funding for the DES Projects has been provided by the U.S. Department of Energy, the U.S. National Science Foundation, the Ministry of Science and Education of Spain, the Science and Technology Facilities Council of the United Kingdom, the Higher Education Funding Council for England, the National Center for Supercomputing Applications at the University of Illinois at Urbana-Champaign, the Kavli Institute of Cosmological Physics at the University of Chicago, the Center for Cosmology and Astro-Particle Physics at the Ohio State University, the Mitchell Institute for Fundamental Physics and Astronomy at Texas A\&M University, Financiadora de Estudos e Projetos, Funda{\c c}{\~a}o Carlos Chagas Filho de Amparo {\`a} Pesquisa do Estado do Rio de Janeiro, Conselho Nacional de Desenvolvimento Cient{\'i}fico e Tecnol{\'o}gico and the Minist{\'e}rio da Ci{\^e}ncia, Tecnologia e Inova{\c c}{\~a}o, the Deutsche Forschungsgemeinschaft, and the Collaborating Institutions in the Dark Energy Survey.
The Collaborating Institutions are Argonne National Laboratory, the University of California at Santa Cruz, the University of Cambridge, Centro de Investigaciones Energ{\'e}ticas, Medioambientales y Tecnol{\'o}gicas-Madrid, the University of Chicago, University College London, the DES-Brazil Consortium, the University of Edinburgh, the Eidgen{\"o}ssische Technische Hochschule (ETH) Z{\"u}rich,  Fermi National Accelerator Laboratory, the University of Illinois at Urbana-Champaign, the Institut de Ci{\`e}ncies de l'Espai (IEEC/CSIC), the Institut de F{\'i}sica d'Altes Energies, Lawrence Berkeley National Laboratory, the Ludwig-Maximilians Universit{\"a}t M{\"u}nchen and the associated Excellence Cluster Universe, the University of Michigan, the National Optical Astronomy Observatory, the University of Nottingham, The Ohio State University, the OzDES Membership Consortium, the University of Pennsylvania, the University of Portsmouth, SLAC National Accelerator Laboratory, Stanford University, the University of Sussex, and Texas A\&M University. Based in part on observations at Cerro Tololo Inter-American Observatory, National Optical Astronomy Observatory, which is operated by the Association of Universities for Research in Astronomy (AURA) under a cooperative agreement with the National Science Foundation.

%DECALS
The Legacy Surveys consist of three individual and complementary projects: the Dark Energy Camera Legacy Survey (DECaLS; Proposal ID \#2014B-0404; PIs: David Schlegel and Arjun Dey), the Beijing-Arizona Sky Survey (BASS; NOAO Prop. ID \#2015A-0801; PIs: Zhou Xu and Xiaohui Fan), and the Mayall z-band Legacy Survey (MzLS; Prop. ID \#2016A-0453; PI: Arjun Dey). DECaLS, BASS and MzLS together include data obtained, respectively, at the Blanco telescope, Cerro Tololo Inter-American Observatory, NSF’s NOIRLab; the Bok telescope, Steward Observatory, University of Arizona; and the Mayall telescope, Kitt Peak National Observatory, NOIRLab. Pipeline processing and analyses of the data were supported by NOIRLab and the Lawrence Berkeley National Laboratory (LBNL). The Legacy Surveys project is honored to be permitted to conduct astronomical research on Iolkam Du’ag (Kitt Peak), a mountain with particular significance to the Tohono O’odham Nation.

NOIRLab is operated by the Association of Universities for Research in Astronomy (AURA) under a cooperative agreement with the National Science Foundation. LBNL is managed by the Regents of the University of California under contract to the U.S. Department of Energy.

This project used data obtained with the Dark Energy Camera (DECam), which was constructed by the Dark Energy Survey (DES) collaboration. Funding for the DES Projects has been provided by the U.S. Department of Energy, the U.S. National Science Foundation, the Ministry of Science and Education of Spain, the Science and Technology Facilities Council of the United Kingdom, the Higher Education Funding Council for England, the National Center for Supercomputing Applications at the University of Illinois at Urbana-Champaign, the Kavli Institute of Cosmological Physics at the University of Chicago, Center for Cosmology and Astro-Particle Physics at the Ohio State University, the Mitchell Institute for Fundamental Physics and Astronomy at Texas A\&M University, Financiadora de Estudos e Projetos, Fundacao Carlos Chagas Filho de Amparo, Financiadora de Estudos e Projetos, Fundacao Carlos Chagas Filho de Amparo a Pesquisa do Estado do Rio de Janeiro, Conselho Nacional de Desenvolvimento Cientifico e Tecnologico and the Ministerio da Ciencia, Tecnologia e Inovacao, the Deutsche Forschungsgemeinschaft and the Collaborating Institutions in the Dark Energy Survey. The Collaborating Institutions are Argonne National Laboratory, the University of California at Santa Cruz, the University of Cambridge, Centro de Investigaciones Energeticas, Medioambientales y Tecnologicas-Madrid, the University of Chicago, University College London, the DES-Brazil Consortium, the University of Edinburgh, the Eidgenossische Technische Hochschule (ETH) Zurich, Fermi National Accelerator Laboratory, the University of Illinois at Urbana-Champaign, the Institut de Ciencies de l’Espai (IEEC/CSIC), the Institut de Fisica d’Altes Energies, Lawrence Berkeley National Laboratory, the Ludwig Maximilians Universitat Munchen and the associated Excellence Cluster Universe, the University of Michigan, NSF’s NOIRLab, the University of Nottingham, the Ohio State University, the University of Pennsylvania, the University of Portsmouth, SLAC National Accelerator Laboratory, Stanford University, the University of Sussex, and Texas A\&M University.

BASS is a key project of the Telescope Access Program (TAP), which has been funded by the National Astronomical Observatories of China, the Chinese Academy of Sciences (the Strategic Priority Research Program “The Emergence of Cosmological Structures” Grant \# XDB09000000), and the Special Fund for Astronomy from the Ministry of Finance. The BASS is also supported by the External Cooperation Program of Chinese Academy of Sciences (Grant \# 114A11KYSB20160057), and Chinese National Natural Science Foundation (Grant \# 12120101003, \# 11433005).

The Legacy Survey team makes use of data products from the Near-Earth Object Wide-field Infrared Survey Explorer (NEOWISE), which is a project of the Jet Propulsion Laboratory/California Institute of Technology. NEOWISE is funded by the National Aeronautics and Space Administration.

The Legacy Surveys imaging of the DESI footprint is supported by the Director, Office of Science, Office of High Energy Physics of the U.S. Department of Energy under Contract No. DE-AC02-05CH1123, by the National Energy Research Scientific Computing Center, a DOE Office of Science User Facility under the same contract; and by the U.S. National Science Foundation, Division of Astronomical Sciences under Contract No. AST-0950945 to NOAO.

The Photometric Redshifts for the Legacy Surveys (PRLS) catalog used in this paper was produced thanks to funding from the U.S. Department of Energy Office of Science, Office of High Energy Physics via grant DE-SC0007914.

%%%%%%%%%%%%%%%%%%%%%%%%%%%%%%%%%%%%%%%%%%%%%%%%%%
\section*{Data Availability}
MGCLS products are publicly available (\url{https://doi.org/10.48479/7epd-w356}). Reprocessed products created as part of this work are available on reasonable request from the authors.



\bibliographystyle{mnras}
\bibliography{refs} % if your bibtex file is called example.bib

%%%%%%%%%%%%%%%%%%%%%%%%%%%%%%%%%%%%%%%%%%%%%%%%%%

%%%%%%%%%%%%%%%%% APPENDICES %%%%%%%%%%%%%%%%%%%%%

%%%%%%%%%%%%%%%%%%%%%%%%%%%%%%%%%%%%%%%%%%%%%%%%%%

% Don't change these lines
\bsp	% typesetting comment
\label{lastpage}
\end{document}